\documentclass[aps,twocolumn,twoside,nofootinbib,prl,preprintnumbers,superscriptaddress]{revtex4}
\usepackage{graphicx}
 
\def\lsim{\mathrel{\raise.3ex\hbox{$<$\kern-.75em\lower1ex\hbox{$\sim$}}}} 
\def\gsim{\mathrel{\raise.3ex\hbox{$>$\kern-.75em\lower1ex\hbox{$\sim$}}}}

\begin{document} 

\bibliographystyle{revtex}

\preprint{hep-ph/0406026}

\pagestyle{plain}
 
\title{Kaluza-Klein Dark Matter and the Positron Excess} 

\author{Dan Hooper}
\affiliation{\mbox{Astrophysics, University of Oxford, Oxford, UK}}

\author{Graham D. Kribs} 
\affiliation{\mbox{School of Natural Sciences, 
Institute for Advanced Study, Princeton, NJ 08540, USA}}

\begin{abstract}

The excess of cosmic positrons observed by the HEAT experiment may be
the result of Kaluza-Klein dark matter annihilating in the galactic halo.
Kaluza-Klein dark matter annihilates dominantly into charged leptons 
that yield a large number and 
hard spectrum of positrons per annihilation.  Given a Kaluza-Klein dark 
matter particle with a mass in the range of 300-400 GeV, no exceptional 
substructure or clumping is needed in the local distribution of dark matter 
to generate a positron flux that explains the HEAT observations. 
This is in contrast to supersymmetric dark matter that requires 
unnaturally large amounts of dark substructure to produce the 
observed positron excess. 
Future astrophysical and collider tests are outlined that will confirm
or rule out this explanation of the HEAT data.

\end{abstract}


\maketitle 


In 1994 and 1995, the High-Energy Antimatter Telescope (HEAT) reported 
an excess of cosmic positrons, peaking in the range of 7-10 GeV, 
and continuing to higher energies \cite{heat1995}.  In 2000, 
an additional HEAT flight confirmed this observation \cite{heat2000}. 
Many previous experiments, although less precise, also recorded a larger 
than expected positron flux above about 10 GeV (see Ref.~\cite{heat1995} 
and references therein).  The study of the astrophysical production 
of positrons \cite{Protheroe} has been thoroughly investigated \cite{secbg}, 
with the conclusion that the ratio of cosmic positrons to electrons 
above about 10 GeV is higher than is suggested by secondary production 
in a model of a diffusive halo.

Galactic positrons potentially provide an interesting probe of particle 
dark matter annihilation in the galactic halo \cite{early}.  
The prospects for supersymmetric dark matter annihilation producing 
positrons, including within the context of the HEAT observations, 
have been extensively discussed, for example, in 
Refs.~\cite{early,morerecent,posbaltz,hooperpos,deboer}. 
In supersymmetric models, the lightest supersymmetric particle 
is stable (with exact $R$-parity) and is usually a neutralino.
Neutralino annihilation directly to $\ell^+\ell^-$ is, however, 
helicity-suppressed, and thus positrons arise only through cascade 
decays such as from decays of gauge bosons.  
This typically results in a rather soft spectrum of positrons and is, 
therefore, hard to reconcile with
the positron flux and spectrum observed by HEAT without an unnaturally 
large degree of clumpiness in our galactic neighborhood.

A fascinating alternative to supersymmetric dark matter arises in 
models with ``universal'' extra dimensions \cite{ew300}.
The premise is that all standard model fields propagate in a higher 
dimensional bulk that is compactified on a space whose size is 
about TeV$^{-1}$ (for earlier work, see \cite{kkdark,universal}).
Higher dimensional momentum conservation in the bulk translates
into Kaluza-Klein (KK) mode number conservation in four dimensions
that is broken by orbifold boundary conditions to a discrete subgroup,
called KK parity.  All odd-level KK modes are odd under KK parity, 
and therefore the lightest level-one KK particle (LKP) does not decay.  
The most natural candidate for the LKP is the first KK excitation of the 
hypercharge gauge boson, $B^{(1)}$ \cite{universal,radiative,kkuniversal,tait}.
In addition to being stable, neutral and colorless, the thermal relic 
density of $B^{(1)}$s is consistent with the measurements from WMAP 
when the mass of $B^{(1)}$ is in the range of hundreds of GeV up to 
about a TeV \cite{tait}.  The precise LKP relic density depends on the 
mass spectrum of the level-one KK excitations, however.  
We will consider LKPs with masses as light as allowed by precision 
electroweak constraints, $m_{B^{(1)}} \gsim 300$ GeV 
\cite{ew300}.

Direct and indirect detection strategies for $B^{(1)}$ Kaluza-Klein 
Dark Matter (KKDM) have been 
explored \cite{feng,kkindirect1,direct,kkindirect2}.
Indirect detection is particularly promising given the large dark matter
mass, annihilation cross section and
annihilation fraction into leptons \cite{feng,kkindirect1}.
The positron flux and spectrum from annihilating KKDM was first
considered in Ref.~\cite{feng}.  In this letter, we revisit the positron flux 
and spectrum specifically to explore the possibility that KKDM annihilating 
in the galactic halo is responsible for the positron excess observed by HEAT.

$B^{(1)}$ dark matter dominantly annihilates through $t$-channel exchange 
of other level-one KK particles.  The annihilation cross section into 
fermions is proportional to the final state fermion's hypercharge to 
the fourth power. Thus right-handed
leptons dominate with an annihilation fraction of $20$-$23\%$ 
per generation for an approximately degenerate level-one KK spectrum.  
This means energetic (``hard'') positrons are copiously produced in 
KKDM annihilation both directly and through cascades of muon and 
tau decay.  The remaining annihilation fraction is primarily into 
right-handed up-type quarks that can also produce positrons via 
cascading, but mainly at lower energies. 

\begin{figure}[t]
\centerline{\includegraphics[width=0.8\hsize]{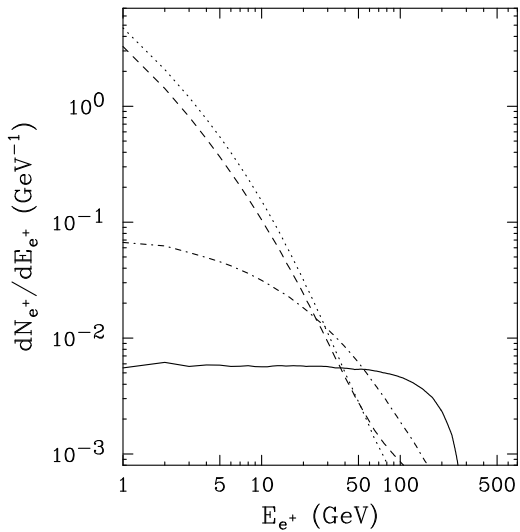}}
\caption{The positron spectrum from generic particle dark matter 
annihilations, prior to propagation, for selected annihilation modes with
$m_{\rm DM} = 300$ GeV\@.  
Solid, dot-dash, dotted and dashed lines correspond to the positron spectrum 
per annihilation into $\mu^+ \mu^-$, $\tau^+\tau^-$, $b \bar{b}$ and 
gauge bosons, respectively.  Charged lepton final states clearly produce 
a considerably harder spectrum of positrons than in other modes. 
The spectrum for annihilation into $e^+ e^-$ (not shown) is trivially 
a delta function at an energy equal to the dark matter particle mass.}
\label{spectrum}
\end{figure}

In Fig.~\ref{spectrum}, we show the positron spectrum that results 
from generic particle dark matter annihilation to $\tau^+ \tau^-$, 
$\mu^+ \mu^-$, $b \bar{b}$ and gauge boson pairs.  Clearly, annihilation 
into $e^+e^-$, $\mu^+\mu^-$ and $\tau^+\tau^-$ produces a much harder 
spectrum than the modes that typically dominate for supersymmetric dark 
matter (gauge bosons or $b \bar{b}$).  These initial positron spectra, 
including cascade decays, were calculated using PYTHIA \cite{pythia} 
as it is implemented in the DarkSusy package \cite{darksusy}.

Following their production, positrons travel through the galactic halo 
under the influence of interstellar magnetic fields and lose energy 
via inverse Compton and synchrotron processes. The effects of propagation 
on the positron spectrum can be calculated using a standard 
diffusion model \cite{posbaltz,MS99}.
Such a technique is limited, however, by the uncertainties in the 
relevant parameters, such as the diffusion constant and energy loss rate. 

Cosmic ray measurements (primarily the boron to carbon ratio) 
indicate a diffusion constant best fit to 
$K(E_{e^{+}}) = 3.3 \times 10^{28} (E_{e^{+}}/1\,\rm{GeV})^{0.47} \, 
\rm{cm}^2/\rm{s}$ \cite{strong02} 
with 20 to 25$\%$ uncertainties at the 1$\sigma$ confidence level. 
For the positron energy loss rate, only a rough estimate is possible, 
and the value of this parameter could vary with location.  We use a value 
for the energy loss rate of 
$b(E_{e^{+}}) = 10^{-16} (E_{e^{+}}/1\,\rm{GeV})^2 \,\, \rm{GeV/s}$. 
We consider a $2 L =8\,$ kpc thick slab for the diffusion zone, 
which is the width best fit to observations \cite{strong02,L}. 
While we have used a modified isothermal sphere profile, we find that 
other profiles such as NFW produce very similar results. The effect of 
varying $L$ is also small. This is because positrons, 
unlike gamma-rays and anti-protons, travel only a few kpc before 
losing their energy.  For further discussion of two-zone diffusion models, 
see Refs.~\cite{posbaltz,L,2zonediffusion}.

To minimize the effects of solar modulation, the spectrum of cosmic positrons 
is generally shown as a ``positron fraction'', or the ratio of positrons 
to positrons plus electrons at a given energy. We convert our positron flux 
to a positron fraction by using the spectrum of secondary positrons, 
secondary electrons and primary electrons found in Ref.~\cite{secbg}. 
This flux (without a dark matter contribution) constitutes the 
background to a potential signal.

\begin{figure}[t]
\centerline{\includegraphics[width=0.9\hsize]{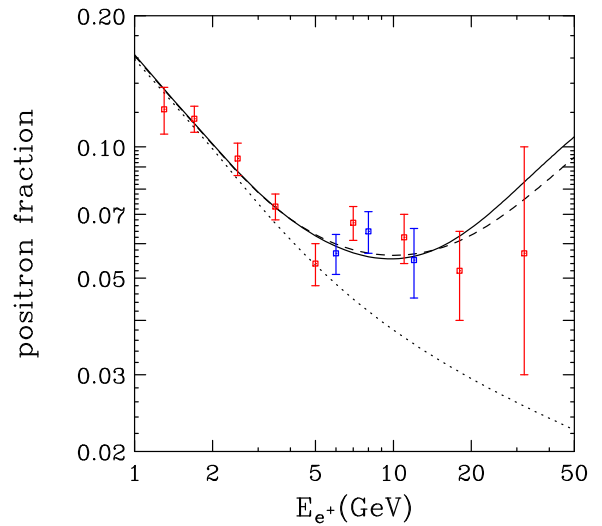}}
\caption{The positron fraction from annihilation of KKDM is shown as 
a function of positron energy.  The solid and dashed lines represent 
300 and 600 GeV $B^{(1)}$s, respectively. The annihilation rate was treated 
as a free parameter, used for normalization. The dotted line represents 
the background predicted with no contribution from dark matter annihilation. 
The error bars shown are from the 1994-95 and 2000 HEAT flights. 
The propagation parameters 
$K(E_{e^{+}}) = 3.3 \times 10^{28} (E_{e^{+}}/1\,\rm{GeV})^{0.47} \, 
\rm{cm}^2/\rm{s}$, 
$b(E_{e^{+}}) = 10^{-16} (E_{e^{+}}/1\,\rm{GeV})^2 \,\, \rm{GeV/s}$ and 
$L=4\,$kpc were used.}
\label{posfrac1}
\end{figure}

The positron fraction predicted from KKDM annihilation is shown as
a function of positron energy in Fig.~\ref{posfrac1}.  The level-one
KK spectrum was assumed to be almost degenerate 
although we found only very slight variation for a spectrum including 
the effects of radiative corrections~\cite{radiative}. Comparing our 
results to the measurements of the 1994-95 and 2000 HEAT flights, 
it is clear that above 7-8 GeV the background-only curve fails to 
match the data while KKDM annihilation can provide a reasonably good 
fit to the data.

\begin{figure}[t]
\centerline{\includegraphics[width=0.9\hsize]{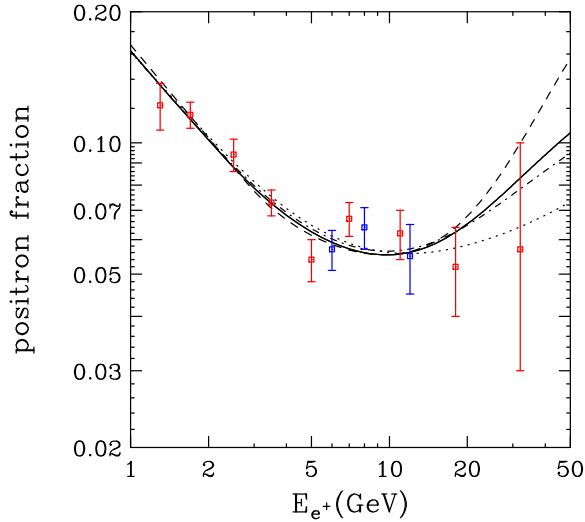}}
\caption{The positron fraction from annihilation of KKDM for several 
choices of propagation parameters. The solid line represents the model 
with the same propagation parameters as in Fig.~\ref{posfrac1}.  
The dashed line is for a model with an energy loss rate smaller by a 
factor of two.  The models represented by dot-dashed and dotted lines 
use the full energy loss rate but diffusion constants that are 
$80\%$ and $50\%$ of the value used in Fig.~\ref{posfrac1}.  Lastly, 
the spectrum with \emph{both} half the diffusion constant and half the 
energy loss rate falls almost exactly on top of the solid line. 
For all cases, $m_{B^{(1)}} = 300$ GeV and $L=4$~kpc were used.}
\label{posfrac2}
\end{figure}

With substantial uncertainties in the propagation parameters, 
it is important to consider the effect of varying these quantities 
on the positron spectrum.  In Fig.~\ref{posfrac2}, we show the 
positron fraction for $m_{B^{(1)}} = 300$ GeV with various choices 
of the diffusion constant and energy loss rate. 

To compare our propagation model and parameters with those used in 
other studies, we remark on two other collaborations' treatment 
of this problem. First, Edsj\"o and Baltz \cite{posbaltz} used a 
considerably lower diffusion constant with a stronger energy dependence: 
$K(E_{e^+}) \propto E_{e^+}^{0.6}$. This was also used in the defaults 
of the DarkSusy package \cite{darksusy}. Alternatively, the more recent 
work by de Boer \emph{et al.} uses a larger diffusion constant with a 
weaker energy dependence, 
$K(E_{e^{+}}) = 4.2 \times 10^{28} (E_{e^{+}}/1\,\rm{GeV})^{0.33} \, 
\rm{cm}^2/\rm{s}$, and a smaller energy loss rate 
$b(E_{e^{+}}) = 5 \times 10^{-17} (E_{e^{+}}/1\,\rm{GeV})^2 \,\, \rm{GeV/s}$ 
\cite{deboer}.  The net effect of these choices is a considerably 
harder spectrum, which helps to explain why they could find reasonable 
fits to the HEAT data using supersymmetric dark matter that annihilates 
largely to $b \bar{b}$.  Finally, Ref.~\cite{MS99} provides a set of 
Green's functions for the propagation of positrons while fitting several 
sources of astrophysical data.  
We find that this technique gives a very similar spectrum.

From Figs.~\ref{posfrac1} and \ref{posfrac2} we see that KKDM clearly fits 
the data considerably better than the background.
In fact, for all of the variations of the parameters that we have considered, 
we consistently found a good fit to the data, up to normalization. 
In Table~\ref{param-table}, we show the $\chi^2$ 
per degree of freedom of the fit to the data for several $B^{(1)}$ masses 
and propagation models.  It is remarkable that KKDM is able to fit the 
spectral shape of the HEAT observations to better than $\chi^2$=1.1 
per degree of freedom in all of these cases.

The annihilation cross section of KKDM into fermions in the low velocity limit 
is $\langle \sigma v \rangle = 95 g^4_1 / 324 \pi m^2_{B^{(1)}}$ \cite{tait}. 
This is about 7 pb for $m_{B^{(1)}} = 300$ GeV\@.  Using this cross section 
and a smooth dark matter halo profile (without clumps), the flux 
of positrons produced in dark matter annihilations can be calculated.  
Spatial density variations from a 
smooth distribution of dark matter are expected to enhance the 
effective annihilation rate by a factor of, perhaps, 2 to 4, but not 
much more \cite{hooperpos}.  This astrophysical increase in the rate 
is commonly called the ``boost factor''.  

\begin{table}[t]
\begin{tabular}{l|cccc}
\hline 
Model & $\chi^2/d.o.f.$  & BF$_{\rho=0.3}$ & BF$_{\rho=0.8}$ & \\
\hline \hline
$m=300$, $K_0=3.3$, $b_0=1\,\,\,$ & 10.8/12  & 24.1 & 3.4 & \\
$m=400$, $K_0=3.3$, $b_0=1\,\,\,$ & 10.1/12  & 66.7 & 9.4 & \\
$m=500$, $K_0=3.3$, $b_0=1\,\,\,$ & 9.7/12  & 139.3 & 19.6 & \\
$m=600$, $K_0=3.3$, $b_0=1\,\,\,$ & 9.4/12  & 253.8 & 35.7 & \\
\hline 
$m=300$, $K_0=3.3$, $b_0=0.5\,\,\,$ & 12.9/12  & 23.8 & 3.3 & \\
$m=300$, $K_0=2.6$, $b_0=1\,\,\,$ & 10.2/12  & 21.6 & 3.0 & \\
$m=300$, $K_0=1.7$, $b_0=1\,\,\,$ & 10.2/12  & 16.7 & 2.3 & \\
$m=300$, $K_0=1.7$, $b_0=0.5\,\,\,$ & 10.8/12  & 12.1 & 1.7 & \\
\hline 
$m=400$, $K_0=3.3$, $b_0=0.5\,\,\,$ & 11.7/12  & 59.9 & 8.4 & \\
$m=400$, $K_0=2.6$, $b_0=1  \,\,\,$ & 9.8/12  & 56.3 & 7.9 & \\
$m=400$, $K_0=1.7$, $b_0=1  \,\,\,$ & 10.2/12  & 44.2 & 6.2 & \\
$m=400$, $K_0=1.7$, $b_0=0.5\,\,\,$ & 10.1/12  & 33.4 & 4.7 & \\
\hline 
\end{tabular}
\caption{The quality of the spectral fit ($\chi^2$ per degree of freedom) 
and the boost factors required for various $B^{(1)}$ masses and propagation 
parameters. $m$ is the mass of $B^{(1)}$ in GeV\@. $K_0$ is the 
diffusion constant in units of 
$10^{28} (E_{e^{+}}/1\,\rm{GeV})^{0.47} \, \rm{cm}^2/\rm{s}$. 
$b_0$ is the positron energy loss rate in units of 
$10^{-16} (E_{e^{+}}/1\,\rm{GeV})^2 \,\, \rm{GeV/s}$. 
The columns BF$_{\rho=0.3}$ and BF$_{\rho=0.8}$ contain the boost factors 
required assuming a local dark matter density of $\rho=0.3$ and 
$\rho=0.8 \, \rm{GeV/cm}^3$, respectively.}
\label{param-table}
\end{table}

In Table~\ref{param-table}, the boost factors needed for KKDM to fit
the data are shown for various $B^{(1)}$ masses and propagation parameters. 
The last two columns correspond to the boost factors that would be needed 
given a local dark matter density of $\rho=0.3$ and $0.8 \, \rm{GeV/cm}^3$, 
respectively.  The first value is the best fit density, 
while the second value is approximately the largest density consistent 
with observations for a reasonable halo profile \cite{BUB}. 
For a light LKP ($m_{B^{(1)}} = 300$ GeV), we find that the boost factor 
required is in the range of $12$-$24$ for the best fit local density. 
Given a higher local density ($0.8 \, \rm{GeV/cm}^3$), the boost factor 
required is
in the range ($1.7$-$3.4$) that is well within astrophysical expectations
for local dark matter clumpiness.  Heavier $B^{(1)}$s require larger
boost factors as illustrated for $m_{B^{(1)}} = 400$ GeV\@. 
Note that it is possible to reduce the boost factor by up to about a 
factor of two at the expense of worsening the fit to the HEAT observations.
For a LKP heavier than about 400 GeV, the positron flux produced 
is likely to be too small to account for the HEAT observations without an
unnaturally large degree of dark substructure.

In Ref.~\cite{tait} the authors found that the relic density of $B^{(1)}$ 
KKDM falls with the range measured by WMAP for $m_{B^{(1)}} = 550$ to 800 
GeV\@.  The lower value of the 
$B^{(1)}$ mass corresponds to a level-one KK spectrum with right-handed 
KK leptons only 1\% heavier than $B^{(1)}$, leading to significant 
coannihilations.  To naturally fit the HEAT observations we ideally need 
$m_{B^{(1)}} \lsim 400$ GeV, and therefore, the $B^{(1)}$ relic density 
is naively too low by a factor of $2$-$3$.  However, variations in the KK 
spectrum, such as lowering the masses of the KK quarks, leads to additional
coannihilation channels (which were not calculated in Ref.~\cite{tait}) 
that can enhance the relic density and therefore lower the mass 
of $B^{(1)}$ needed to get the relic density up into the WMAP range.  
There could also be 
nonthermal sources of KKDM that boost the relic density.  
In any case, we are encouraged that the thermal relic density is at 
least roughly in the right range that is consistent with the $B^{(1)}$ 
mass range needed to explain the HEAT observations.

Future experiments, such as PAMELA and AMS-02, will be capable of 
measuring the cosmic positron spectrum to much higher energies and with 
greater precision.  The confirmation in such experiments of a rise in 
the positron spectrum, as shown in Figs.~\ref{posfrac1} and~\ref{posfrac2}, 
would further favor KKDM as the source of the positron excess. 
In addition, a distinctive spike in the spectrum at $E_{e^+} = m_{B^{(1)}}$ 
is expected \cite{feng} that would provide a good measurement of
the $B^{(1)}$ mass.  Indirect detection is also expected at next generation 
neutrino telescopes such as IceCube that should find between tens and a 
thousand events per year from KKDM annihilations in the Sun 
\cite{kkindirect1}.  Existing neutrino telescopes, such as AMANDA-II, 
may potentially be sensitive to this scenario as well.  
Direct dark matter detection experiments are also approaching 
the sensitivity needed to observe such a WIMP \cite{direct,feng}.  
Finally, future collider experiments including the Tevatron and 
particularly the Large Hadron Collider (LHC) are expected to produce 
most of the level-one and possibly higher level KK modes that will
lead to fascinating (and confusing) signals \cite{kkuniversal} 
in this scenario.

In conclusion, 
we have shown that Kaluza-Klein dark matter annihilating in the 
galactic halo can account for the excess in the cosmic positron spectrum 
observed by the HEAT experiment.  We find an excellent fit to the 
observed positron spectrum for a wide range of positron propagation 
parameters and $B^{(1)}$ dark matter masses.  
Additionally, with reasonable values 
of the local dark matter density and clumpiness, the annihilation rate 
of 300-400 GeV Kaluza-Klein dark matter and the corresponding positron 
flux can be sufficient to account for the HEAT observations.  
This is in contrast to supersymmetric dark matter, in which an unnatural 
amount of dark matter substructure is invariably required to produce 
the necessary positron flux.  If the HEAT observations are confirmed by 
future measurements of a rising cosmic positron spectrum from PAMELA 
and AMS-02, then interpreting the HEAT excess as arising from KKDM
annihilation in the galactic halo implies
other dark matter detection and collider experiments should soon see 
confirming signals of a world with extra spatial dimensions only slightly
above the electroweak scale. \\
\indent \emph{Acknowledgments:}
We thank Andrew Strong and Joakim Edsj\"o for useful 
communications. DH is supported by the Leverhulme Trust.
GDK is supported by a Frank and Peggy Taplin Membership and
by the Department of Energy under contract DE-FG02-90ER40542.

\vskip -0.4cm


\end{document}